# Experimental Realization of Higher-order Topological Water Wave Crystal

Jiayin Zhao,[1,*] Yifan Liu,[1,*] Jidong Song,[1] Houqi Ai,[1] Qinhui Jiang,[1] Manjie Li,[1,†] and Mengyao Li[1,‡]

[1] Shenzhen International Graduate School, Tsinghua University, Shenzhen 518055, China

**ABSTRACT**. Free-surface water waves provide a macroscopic platform in which topological localization can be measured through surface-elevation fields. Higher-order topological phases offer a route to localizing wave modes at structural corners, yet their experimental realization in water-wave systems remains largely unexplored. Here, we experimentally demonstrate a water-wave higher-order topological crystal, where surface waves are confined to structural corners and spectrally identified through measured free-surface responses. Bulk band analysis and finite-system simulations predict topological edge and corner states in the designed structure, while frequency-resolved surface-elevation measurements confirm the corresponding localized modes. The measured spectra and field distributions reveal pronounced in-gap corner localization consistent with the predicted higher-order topological modes. These results establish surface water waves as a directly measurable platform for higher-order topological physics, enabling spatial localization and spectral characterization of hydrodynamic wave modes.

## I. INTRODUCTION.

Topological physics provides a framework for understanding and constructing robust boundary and localized modes. Beyond conventional first-order topological phases, higher-order topological insulators (HOTIs) enrich bulk-boundary correspondence by supporting lower-dimensional localized modes, such as corner states in two-dimensional systems [1–5]. Owing to the wave nature of their governing equations, topological principles have been extended beyond electronic systems to a variety of classical-wave platforms, enabling robust transport and localized states in artificial photonic [6–10] and acoustic structures [11–14]. In this context, the two-dimensional Su-Schrieffer-Heeger (SSH) model provides a tractable framework for engineering higher-order topological phases through modulated intra-cell and inter-cell couplings. Although such corner states have been observed and used for functional wave phenomena in several classical-wave systems[15,16], their realization in free-surface water waves remains largely unexplored.

In natural and engineered water environments, such as marine and coastal systems, controlling the propagation and spatial distribution of surface waves is important for wave-energy management and coastal protection [17–19]. Artificial water-wave structures have enabled unconventional phenomena, including surface-wave cloaking [20, 21] and unidirectional propagation or transmission [22, 23]. However, conventional designs often depend on carefully tuned geometries or modal resonances, so their ability to guide waves and localize energy can be vulnerable to structural imperfections. Therefore, topological design addresses this challenge by linking wave localization and transport to band-structure properties. Recent studies have introduced SSH-type models into one-dimensional water-wave systems [24], realized topological gravity-capillary waves and guided transport in water-wave crystals [25, 26], and reported vortex and valley-type topological states in free-surface waves [27–30]. Although these works demonstrate the feasibility of topological control in dispersive water-wave systems, most existing demonstrations emphasize either transport along designed paths or individual modal phenomena. Compact water-wave structures that experimentally realize higher-order topological localization, while also supporting wave guiding and spatially resolved identification of localized modes, remain much less explored. Such integration is desirable for artificial surface-wave systems in which wave energy should be routed and concentrated at prescribed regions, for example in future wave-energy collection or nearshore wave-management concepts.

In this Letter, we realize and investigate higher-order corner states and topological edge states in a two-dimensional SSH water wave structure operating in the few-hertz regime. Finite-element simulations reveal the band topology and eigenmode distributions, predicting edge states together with in-gap corner-localized states in the proposed system. Experiments verify the higher-order topological response through frequency-resolved local spectral measurements and wave-field distributions, revealing pronounced corner localization at the predicted in-gap frequencies. These results establish a geometrically tunable platform that integrates edge states and corner localization in a free-surface water-wave system, providing an experimental approach to localized surface-wave confinement and opportunities for controlled surface-wave energy concentration.

## II. THEORETICAL MODEL AND NUMERICAL METHODS

### A. Governing equation of water waves

*These authors contributed equally to this work.
†Contact author: manjie.li@sz.tsinghua.edu.cn
‡Contact author: mengyaoli@sz.tsinghua.edu.cn

We consider linear surface waves with small amplitude in a water wave crystal. After separating the vertical dependence of the three-dimensional velocity potential, the water-wave problem can be reduced to an effective two-dimensional Helmholtz equation for the horizontal velocity-potential amplitude $\psi(\mathbf{r})$ [31]. Assuming harmonic time dependence, the reduced field satisfies

$$(\nabla_{\perp}^2 + k^2)\psi(\mathbf{r}) = 0 \quad (1)$$

where $r = (x, y)$, $\nabla_{\perp}^2$ is the in-plane Laplacian, and $k$ is the water wave wavenumber. Although Eq. (1) is solved in the horizontal plane, the vertical particle motion is not neglected; it is retained through the depth-dependent part of the three-dimensional velocity potential. The corresponding derivation and the expressions for the horizontal and vertical velocity components are provided in Supplementary Material S1. The physical eigenfrequency is obtained from the finite-depth gravity-capillary dispersion relation

$$\omega^2 = \left(gk + \frac{\sigma k^3}{\rho}\right)\tanh(kh) \quad (2)$$

where $g$ , $\rho$ , $\sigma$ , and $h$ denote the gravitational acceleration, fluid density, effective surface-tension coefficient, and water depth, respectively. In the present reduced model, $\sigma_{eff} = 0.17$ N m$^{-1}$ is adopted as an effective parameter for the structured water-wave system [25]. Its physical interpretation and influence on the frequency conversion are further discussed in Supplementary Material S1.

For a periodic crystal, the Bloch condition is imposed as

$$\psi(\mathbf{r} + \mathbf{R}) = \psi(r)e^{-i\mathbf{k_B}\cdot\mathbf{R}} \quad (3)$$

where $\mathbf{R}$ is the lattice translation vector and $\mathbf{k_B}$ is the Bloch wave vector. The scalar water-wave wavenumber k in Eqs. (1) and (2) should be distinguished from the Bloch wave vector $\mathrm{k_B}$, which specifies the phase relation between adjacent unit cells. The Bloch condition in Eq. (3) reduces the periodic system to a unit-cell eigenvalue problem, and the resulting values of $k$ are converted into eigenfrequencies through the dispersion relation in Eq. (2). Details of the water-wave model and numerical implementation are provided in Supplementary Material S1. By engineering the unit-cell geometry and the relative intra-cell and inter-cell couplings, a topological band gap is opened in the nontrivial 2D SSH configuration. Within this gap, one-dimensional edge bands appear under open boundary conditions, and zero-dimensional corner modes emerge in finite structures, in contrast to spatially extended bulk states.

### *B. Band evolution and modal characteristics*

The designed 2D SSH water wave crystal is constructed on a square lattice, consisting of four square cavities interconnected by narrow channels, as illustrated in Fig. 1(a). Within the tight-binding framework, the cavities serve as localized resonant sites, while the connecting channels control the coupling between adjacent nodes. The lattice constant is $a = 4$ cm, with a cavity side length of $d = 4a/15$ and a fixed water depth of $h = 2$ cm. The parameters $d_x$ and $d_y$ denote the center-to-center distances between adjacent cavities along the x and y directions, respectively. Here, $d_x$ and $d_y$ are held constant $a/2$, while the coupling strength is tuned by varying the channel widths within the unit cell. Specifically, $b_1$ and $b_2$ denote the intra-cell and inter-cell channel widths, respectively. The band structure is calculated along the high-symmetry path Γ–X–M–Γ.

The complete four-band spectrum for different unit-cell geometry configurations of the four-cavity unit cell is shown in Fig. 1(b), with emphasis on the lowest band gap relevant to this work. The shrunken configuration, characterized by stronger intracell coupling, corresponds to the trivial phase and satisfies $b_2 = b_1/3 = d/4.5$ , whereas the expanded configuration, characterized by stronger intercell coupling, corresponds to the nontrivial phase and satisfies $b_1 = b_2/3 = d/4.5$. The resulting band gap lies in the frequency range of 3.72–4.18 Hz, and the trivial bands overlap with the nontrivial case outside the gap. In contrast, the band gap closes at $b_1 = b_2 = (d/1.5 + d/4.5)/2$ , corresponding to the critical point between the two phases. The gap closing and reopening indicate a topological phase transition driven by the unit-cell geometry. The corresponding topological properties are further verified by bulk polarization calculations in Supplementary Material S2. Fig. 1(d) displays the supercell band structure constructed from the trivial and nontrivial unit cells, where an edge-state band emerges within the bulk band gap and partially overlaps with the bulk-state frequency range. Representative mode profiles are shown in Fig. 1(c). The edge modes are localized at the interface between the two inequivalent unit cells, whereas the bulk modes extend over the whole structure.

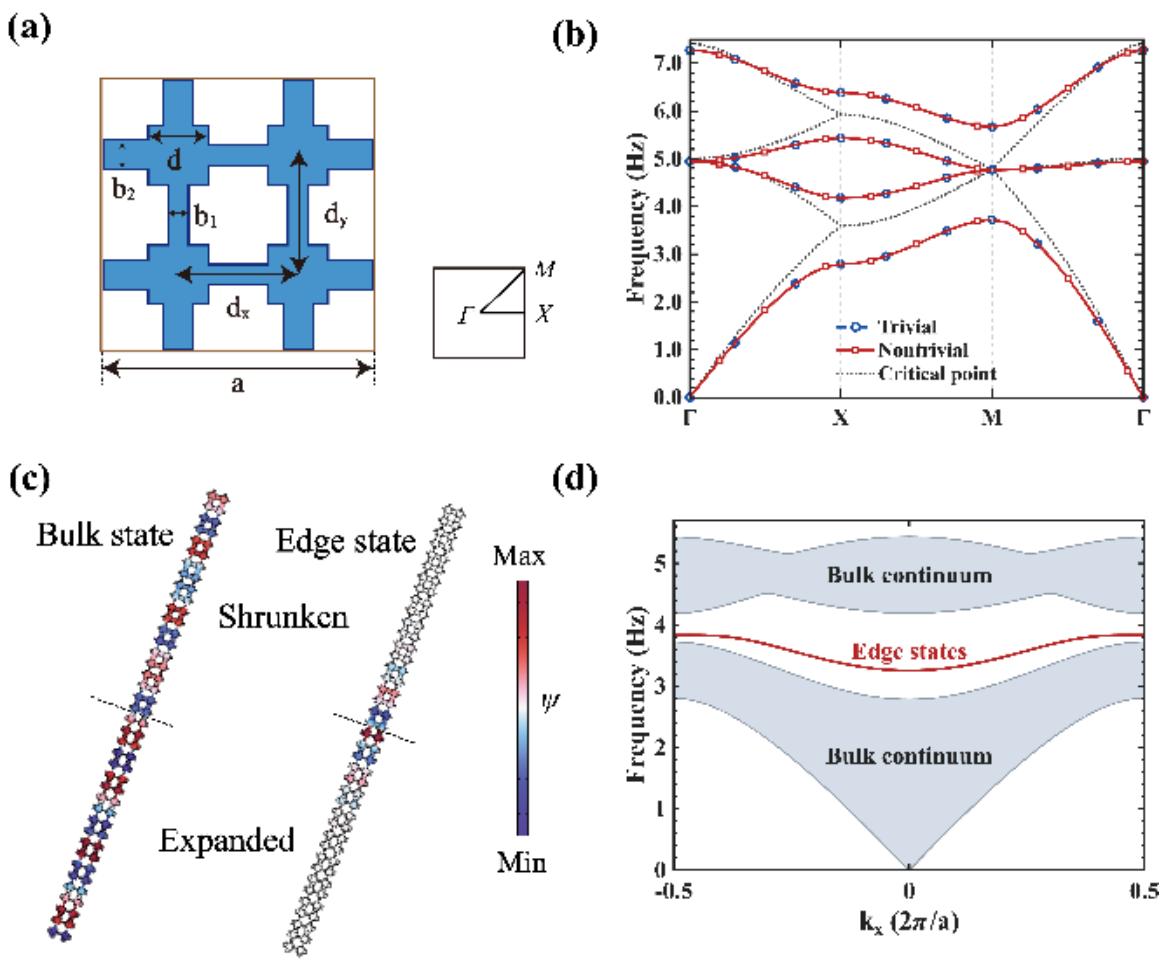

FIG. 1. Band evolution and emergence of edge states in the 2D SSH water wave crystal under different coupling configurations. (a) Schematic of the primitive unit cell in the 2D SSH model. (b) Calculated band structures for different coupling configurations, showing band gap openings in both expanded and shrunken cases. (c) Potential-energy field distributions of bulk and edge states within a supercell composed of expanded and shrunken unit cells. (d) Supercell band structure showing edge-state bands within the band gap.

Based on the band analysis, a finite-sized structure is constructed to further characterize edge and corner states. The water wave crystal is modeled by solving the Helmholtz equation in COMSOL Multiphysics 6.2. The finite structure comprises an 8 × 8 region formed by the nontrivial expanded unit cell shown in Fig. 1(b), surrounded by four layers of the trivial shrunken unit cell, forming a closed interface between the two phases. Fig. 2 shows the simulated eigenmode distributions. Fig. 2(a) presents the eigenfrequency spectrum of the finite channel-coupled 2D SSH water wave crystal, and Figs. 2(b)–2(d) show representative mode profiles at selected frequencies. Within the frequency range of 3.73–3.84 Hz, the modes exhibit pronounced localization along the boundaries, indicating topological edge states. At 3.99 Hz, the field is confined to the corners, showing a higher-order corner state. By contrast, the mode at 4.36 Hz extends over the whole structure, corresponding to a bulk state. These results demonstrate that the proposed channel-coupled 2D SSH water wave crystal supports bulk, edge, and corner states, with the corner state distinctly identified by its spectral position and spatial localization. Additional finite-structure spectra and mode-field identification are provided in Supplementary Material S3.

# III. EXPERIMENTAL VERIFICATION

## *A. Experimental setup and procedures*

We use both qualitative and quantitative approaches to analyze the corner states, edge states, and other features realized by the 2D SSH model in experimental water wave systems. The experimental setup, shown in Fig. 3(a), comprises an acrylic water tank (0.5 m × 0.5 m) serving as the principal domain for wave propagation. A wave generator with manual frequency adjustment and digital display facilitates precise control of the excitation conditions. Illumination and imaging are provided by a light source and high-speed camera, both mounted on a metal frame to ensure image fidelity. Two high-precision distance meters are affixed to the mid-level of the frame and are horizontally translatable to measure water surface displacements at arbitrary locations. Checkerboard sheets placed beneath the tank enable image-based propagation observation.

For practical reasons, including water tank size constraints and the effect of surface tension, the experimental study is restricted to the lower right quadrant of the simulated structure, as shown in Fig. 4(a). This quadrant preserves the corner, edge, and bulk regions required for the experimental verification. Light-curing 3D printing with photopolymer resin is utilized to fabricate the experimental models [32,33].

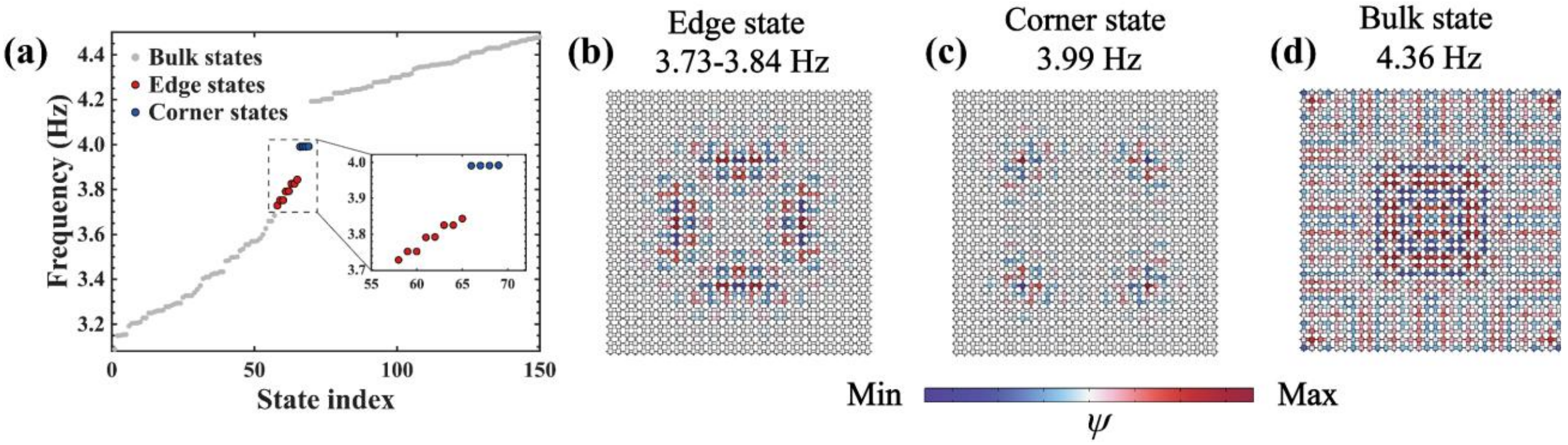

FIG. 2. Bulk, edge, and corner states supported by the 2D SSH water wave crystal in finite sized structures. (a) Numerical results demonstrate the coexistence of bulk, edge, and corner states. (b)–(d) Representative mode profiles and corresponding eigenfrequencies: (b) an in-gap edge state, (c) a topological corner state, and (d) a bulk state.

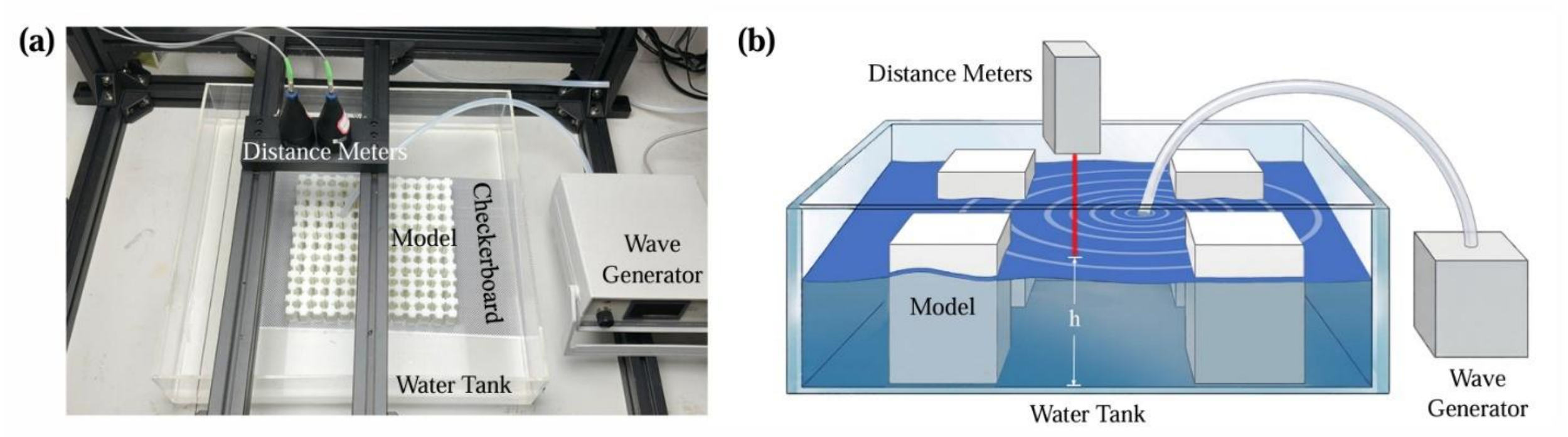

FIG. 3. (a) Overall experimental setup. (b) Schematic diagram of the experimental setup for measuring water wave amplitude using distance meters.

Water-surface elevation was recorded at each selected measurement position using a distance meter. At each excitation setting, the recorded elevation time series was analyzed by fast Fourier transform (FFT) to determine the dominant response frequency [34]. After noise correction, the standard deviation of the elevation time series was used as a measure of the local response amplitude. By systematically varying the excitation frequency, a frequency-amplitude relationship is established for each specific excitation condition. In corner states, energy localization manifests as increased amplitude at targeted locations, thereby confirming the eigenfrequency characteristic of the corner states [35–39].

Moreover, we design a dedicated experiment utilizing image processing methods to verify the global propagation patterns of water waves within the model [25, 40–44]. An initial calibration image is acquired by recording the stationary water surface. Subsequently, a series of images is recorded using the high-speed camera, documenting the displacement of the checkerboard calibration pattern caused by variations in water depth. By employing optical flow algorithms to analyze the horizontal displacements of marked patterns, the spatiotemporal evolution of the water surface is reconstructed, enabling detailed visualization and analysis of wave propagation throughout the experimental structure.

### *B. Experimental observations*

Through amplitude and propagation experiments, we observed the water wave frequency-amplitude relationship and state-specific propagation features of the 2D SSH model. We carried out excitation at locations corresponding to the corner, edge, and bulk states, with the respective positions shown in Fig. 4(a). Additional details about the model size and alternative views are provided in Fig. S9.

For the frequency-response measurements, the nominal excitation frequency on the wave generator was swept from 3.00 to 5.00 Hz in uniform increments of 0.05 Hz. The same acquisition and processing procedure was applied to the corner, edge, and bulk measurements. Using the nominal wave-generator frequency as the horizontal coordinate in Fig. 4(b), the edge- and corner-response maxima occur at setpoints of 3.70 and 3.95 Hz, respectively. FFT analysis of the corresponding surface-elevation time series reveals dominant response frequencies of 3.73 and 3.96 Hz, respectively, as shown in Fig. S10. These experimentally determined eigenfrequencies closely align with simulation results. Within the measurement range, only a single prominent peak is observed for both the edge and the corner state excitations, with no evidence of secondary peaks. For bulk state excitation, the frequency-amplitude curve shows a monotonically increasing trend, and no amplitude peaks are detected within the measurement range. Notably, the amplitude peak associated with edge state excitation is wider than that of the corner state. Given the sensitivity of eigenfrequency to water depth discussed in Supplementary Material S4, the observed experimental deviations remain within an acceptable margin of error.

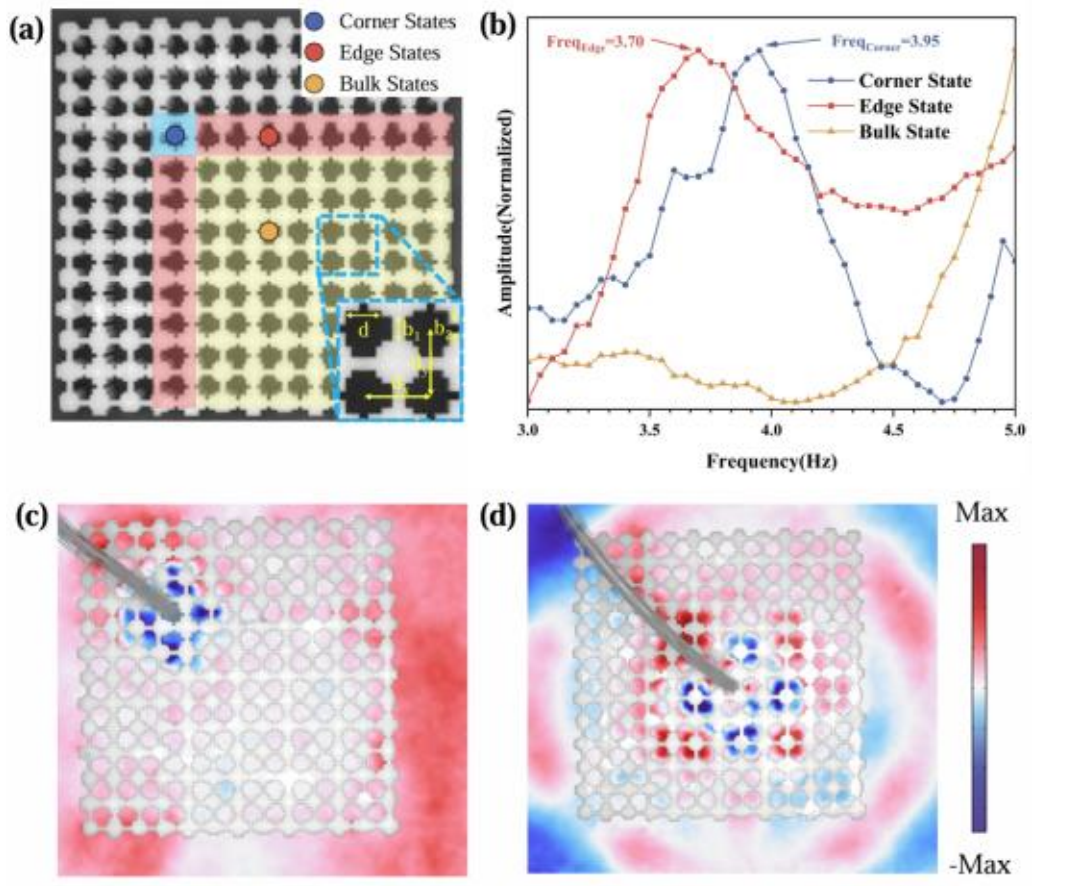

FIG. 4. (a) Experimental realization of the 2D SSH water wave crystal, representing one quadrant of the full simulation model. Corner, edge and bulk state regions are marked with blue, red and yellow transparent rectangles, respectively. (b) Frequency-amplitude relationships for different excitation conditions, measured within the range of 3-5 Hz. (c)-(d) Experimental images of water wave propagation: corner state at 4.00 Hz and bulk state at 3.62 Hz.

In the propagation experiments, we adjusted the excitation frequency to match the eigenfrequencies identified in the amplitude measurements, thereby ensuring selective excitation of the targeted states. For bulk state excitation, we chose a lower frequency of 3.62 Hz to distinguish it from the corner and edge states. Figs. 4(c) and 4(d) present experimental images corresponding to the corner and bulk states, respectively. For the corner state, the water wave energy remains localized near the excitation source, with minimal propagation beyond the immediate region. By contrast, bulk state excitation results in isotropic wave propagation, with energy radiating uniformly throughout the domain. The simulated and experimental water-wave spatial amplitude distributions are compared in Supplementary Material S5. Further field-level analyses are summarized in Supplementary Materials S6–S8, including the edge-state gradient-field magnitude and local orientation, the influence of boundary termination on the edge-state gradient-field distribution, and the robustness of the corner-state mode-field distribution against a point defect.

At the edge-response frequency, time-resolved measurements reveal a wave pattern advancing along the interface. The surface elevation at different positions has different phases, so the interface-guided response is less apparent in a single experimental snapshot than in the full temporal sequence. After the water waves reached a stable state, we conducted continuous imaging and image processing, resulting in Supplemental Movies S1-3. These videos provide a clearer visualization of the unique propagation characteristics of water waves in the edge state and the differences in wave propagation under different excitation states.

These experimental observations demonstrate that the proposed 2D SSH water-wave crystal supports energy localization and boundary-guided propagation, highlighting its potential for advanced wave manipulation in engineered aquatic environments.

## IV. CONCLUSIONS

In conclusion, we have designed and experimentally realized a higher-order topological channel-coupled water wave crystal based on a 2D SSH lattice. The proposed structure supports both topological edge and corner states, demonstrating higher-order topological localization and wave guiding for water waves. The predicted higher-order topological states are consistently verified by numerical simulations and experimental measurements, including wave-amplitude distributions and frequency-resolved local spectral responses that identify the corner-localized modes. Our results provide an experimentally accessible free-surface platform for studying higher-order topological states and a compact route to surface-wave localization and guided transport through topological band engineering. Future research may extend this approach by incorporating tunable structural designs, loss management, and scale adaptation, with potential relevance to water-wave energy localization and harvesting, guided transport, and wave manipulation in ocean engineering and related disciplines.

## ACKNOWLEDGMENTS

This work was supported by the National Key R&D Program of China (Grant No. 2023YFB3811400), Shenzhen Science and Technology Innovation Commission (Grant No. RCBS20231211090547075 and JCYJ20240813112042055), National Natural Science Foundation of China (Grant No. 12404435), Guangdong Science and Technology Commission (Grant No. 2025A1515010774), and Cross-disciplinary Research and Innovation Fund in Tsinghua Shenzhen International Graduate School (Grant No. JC2024005).